\documentstyle[eqsecnum,epsfig,aps,floats,preprint]{revtex}

\newcommand{\beq}{\begin{equation}}
\newcommand{\eeq}{\end{equation}}
\newcommand{\bea}{\begin{eqnarray}}
\newcommand{\eea}{\end{eqnarray}}

\def\laq{~\raise 0.4ex\hbox{$<$}\kern -0.8em\lower 0.62
ex\hbox{$\sim$}~}
\def\gaq{~\raise 0.4ex\hbox{$>$}\kern -0.7em\lower 0.62
ex\hbox{$\sim$}~}

\def \pa {\partial}
\def \ra {\rightarrow}
\def \la {\lambda}
\def \La {\Lambda}
\def \Da {\Delta}
\def \b {\beta}
\def \a {\alpha}

\def \Ga {\Gamma}

\def \sg {\sigma}
\def \da {\delta}
\def \ep {\epsilon}
\def \r {\rho}
\def \om {\omega}
\def \Om {\Omega}

\def \wt {\widetilde}
\def \ep {\epsilon}

\newcommand{\epsi}{e^\psi}
\newcommand{\rhm}{\rho_m}

\begin{document}
\par
\begingroup

\begin{flushright}
BA-TH/01-419\\
CERN-TH/2001-207\\
gr-qc/0108016\\
\end{flushright}
\vskip 0.5true cm

{\large\bf\centering\ignorespaces
Quintessence as a run-away dilaton 
\vskip2.5pt}

{\dimen0=-\prevdepth \advance\dimen0 by23pt
\nointerlineskip \rm\centering
\vrule height\dimen0 width0pt\relax\ignorespaces

M. Gasperini${}^{(1,2)}$, F. Piazza${}^{(3,4)}$ and G.
Veneziano${}^{(4,5)}$ 
\par}
{\small\it\centering\ignorespaces
${}^{(1)}$
Dipartimento di Fisica, Universit\`a di Bari, 
Via G. Amendola 173, 70126 Bari, Italy \\
${}^{(2)}$
Istituto Nazionale di Fisica Nucleare, Sezione di Bari,
Bari, Italy \\
${}^{(3)}$
Dipartimento di Fisica, Universit\`a di Milano Bicocca, \\ 
Piazza delle Scienze 3, I-20126 Milan, Italy \\
${}^{(4)}$
Laboratoire de Physique Th\'eorique, Universit\'e Paris
Sud, 91405 Orsay, France\\
${}^{(5)}$
Theoretical Physics Division, CERN, CH-1211 Geneva 23, Switzerland \\
\par}

\par
\bgroup
\leftskip=0.10753\textwidth \rightskip\leftskip
\dimen0=-\prevdepth \advance\dimen0 by17.5pt \nointerlineskip
\small\vrule width 0pt height\dimen0 \relax

\begin{abstract}
We consider a late-time cosmological model  based on a
recent proposal that the infinite-bare-coupling limit of
superstring/M-theory exists  and has good phenomenological
properties, including a vanishing cosmological constant, and a massless,
decoupled dilaton.  As it runs away to  $+ \infty$, the dilaton can play
the role of  the quintessence field recently advocated to drive the
late-time accelerated expansion of the Universe. If, as suggested by
some string theory examples, appreciable deviations from General
Relativity persist even today in the dark matter sector,  the Universe
may smoothly evolve from an initial  ``focusing" stage, lasting untill
radiation--matter equality, to a  ``dragging" regime, which eventually
gives rise to an accelerated expansion with frozen
$\Omega(\rm{dark\;energy})/\Omega(\rm{dark\;matter})$.  
\end{abstract}

\begin{center}
------------------------------------------------\\

To appear in {\bf Phys. Rev. D}
\end{center}

 \par\egroup
\thispagestyle{plain}
\endgroup

\pacs{}

\section {Introduction}
\label{I}

According to recent astrophysical observations,  our Universe, since a
red shift of ${\cal O}(1)$, appears to have  undergone  a phase of
accelerated expansion \cite{1,2}. This result can be combined with the
recent estimates of the average mass density of the Universe \cite{3},
$\Om_m \simeq 0.3 - 0.4$ (in critical units), and with recent
measurements of the CMB anisotropy peaks \cite{4}, pointing at a nearly
critical total energy density, $ \Om_T \simeq 1$. One is then led to the
conclusion that the present cosmological evolution, when described in
terms of an effective fluid entering  Einstein's equations, should
be (marginally) dominated by a ``dark energy" component $\r_x$
characterized by a (sufficiently) negative effective pressure, $p_x 
< -\r_x/3$.

The simplest candidate for such a missing energy is a positive
cosmological constant $\La$, of order $H_0^2$. Such an identification,
however,  unavoidably raises a series of  difficult questions. In
particular: a)  Why is  $\La$ so small in particle physics units? Explaining
a finite but very small value for $\La$ may turn out to be even harder
than finding a reason why it is exactly zero.  This is the
so-called fine-tuning problem for $\La$,  see for instance
\cite{Weinberg}; and b)  Why  is $\La  \sim  \r_{m 0}$,  where  $\r_{m 0}$
is the  ${\it present}$  value (in Planck units) of the  (dark) matter
energy density? This is the so-called ``cosmic coincidence" problem
\cite{5}.

At present, the most promising scenarios for solving (at least part of)
the above problems introduce a single scalar field, dubbed
``quintessence'' \cite{6}, whose potential goes to zero asymptotically
(leaving therefore just the usual puzzle of why the ``true" cosmological
constant vanishes). The scalar field slowly rolls down such a potential 
reaching infinity (and zero energy) only after an infinite (or very long)
time. While doing so, quintessence  produces an effective,
time-dependent,  cosmic energy density $\r_x$ accompanied by a
sufficiently negative pressure, i.e. a sort of effective cosmological
constant. By making $\La_{eff} \sim H^2$ time-dependent, 
this can naturally  explain
the smallness of the {\it present} effective vacuum energy density. 
However, if, as in General Relativity, dust energy and an effective
cosmological constant have different 
time dependence, it can hardly explain why $\La  \sim  \r_{m 0}$.
For a recent review of the relative merits of a cosmological constant
and quintessence, see Ref. \cite{Binetruy}.

As far as identifying quintessence is concerned, the
inflaton itself could be a  candidate \cite{7}. But also more
exotic possibilities have been considered, in particular
some motivated by the wish to solve the above-mentioned cosmic
coincidence problem \cite{8,9}.  In any case, as  is the case for the
inflaton, the quintessence field does not
have, as yet, an obvious place in any fundamental theory of
elementary particles.  One should also mention, at this point, that, 
 if quintessence may help with the problems typical of the cosmological
constant interpretation,  it is likely to create a new one of its own: in
order to play its  role,  the quintessence field must be extremely light
and can thus mediate a new long--range (of order $H_0^{-1}$) force,
which is strongly  constrained observationally. This is an important
constraint to be imposed on any specific scalar field model
of quintessence, either minimally or non-minimally \cite{nonminimal}
coupled to gravity. 
 
 At first sight, the search for a quintessence candidate in particle
physics looks easier than the one for an inflaton.
For instance, fundamental or effective scalar fields with potentials
running to zero at infinity  are ubiquitous  in supersymmetric field
theories and/or in  string/M-theory. They are usually referred to as 
moduli fields since, in perturbation theory, they parametrize the space
of inequivalent vacua and correspond to exactly flat directions
(equivalently, to exactly massless fields). Non-perturbative effects
(e.g.  gauge-theory instantons) are expected to lift these flat
directions, just preserving those that correspond to small or vanishing
coupling. Examples are the run-away vacua of supersymmetric gauge
theories (see, for instance,  \cite{masiero} for quintessence
 models based on the latter possibility), or the dilaton modulus $\phi$ in
the limit  $\phi \rightarrow - \infty$.
 
However, if we  were to take one of these moduli as 
quintessence, we would immediately run into the problem that the
acceleration of the  Universe should be accompanied by a drift of
interactions towards triviality. For Newton's constant, and even more
so for the fine-structure constant, this kind of time variations
is very strongly constrained. Furthermore, typical couplings of moduli
fields to ordinary matter are of gravitational order, and this creates the
already mentioned problem of new, unwanted long-range
forces.  For all these reasons the conventional attitude towards moduli
fields has been (see e.g. \cite{17}) to postulate that they develop
 non-perturbative potentials, providing  them with both a mass and 
a freezing mechanism (see  however \cite{16} for an alternative that
is closer, in spirit, to the one advocated here).

Another possible problem with the identification of a string modulus
with quintessence is that we would like to freeze the
moduli at values that provide the correct values of the coupling
constant and unification scale of grand unified theories (GUTs). For
instance, the  dilaton and compactification volume $V_6$ 
should be frozen at values such that \cite{12}
\beq
\alpha_{GUT}^{-1} \sim (M_P/M_{s})^2 \sim
e^{-\phi}~, ~~~~~~~M_P/M_{GUT} \sim \alpha_{GUT}^{-2/3} ~g_s^{1/6},
\label{1}
\eeq
where $M_P$, $M_s$ and $M_{GUT}$  are the  Planck, string and GUT
scales, $g_s = e^{\phi} V_6 M_s^6$ is the string coupling,
 and $e^{\phi}$ is the tree-level
effective four-dimensional coupling (thus, in  more standard
string-theory notation \cite{10}, our dilaton is related to the real part
of the  $S$ modulus by ${\rm Re}\{S\}$ = $e^{-\phi}$).

Unfortunately, it looks unlikely that non-perturbative effects
will be significant enough in this region to stabilize the moduli.
 Also, perturbative unification 
gives too low a value for  $M_{P}/M_{GUT}$ \cite{12,Witten}.
In this respect, the situation can be drastically improved by considering
the $M$-theory limit, $g_s \rightarrow \infty$, while still keeping the
four-dimensional effective couplings perturbative ($S \gg 1$)
 \cite{Witten}. Even then,  the moduli would presumably
freeze out in a typical  (and cosmologically tiny) particle-physics time,
and therefore  cannot implement  the conventional, slow-roll
quintessential scenario. In spite of these difficulties,
unconventional models of quintessence  based on the stabilization of
the dilaton in the perturbative regime are not completely excluded, 
as recently discussed by one of us \cite{14}. 

There is, however, another possibility for  making the  dilaton a 
candidate for quintessence. As we have already mentioned, the region
 of large negative $\phi$ corresponds to the trivial vacuum. 
The idea that the Universe may have started,
long before the big bang, in this region  is actually the basis of
the so-called pre-big bang scenario in string cosmology (for recent
reviews see \cite{13a}).  Here we are asking instead whether the
dilaton  can play the role of quintessence at very {\it late} times
(such as
today), not by evolving  towards $- \infty$ and triviality, but by going
towards $+ \infty$ and strong coupling. Such a proposal looks absurd at
first since, if we do  not see a drift towards zero coupling, we do not
experience one  towards increasing strength either. In order for this
idea to make sense we have to assume that the strong-coupling limit of
string/M-theory exists, is smooth, and resembles our world. Can this 
make sense at all? 

It has been argued by one of us \cite{13} that the answer to the
above question can be affirmative, if we assume a certain structure
of the quantum loop corrections to the string effective action suggested
by large-$N$ counting arguments. In the strong coupling limit (which
could either be the self-dual value $S=1$ or, if $S$-duality is broken,
$\phi \rightarrow + \infty$) gravitational and gauge coupling would be
determined entirely by loop corrections (as in the so-called
induced-gravity/gauge idea \cite{Sakharov}), and would ``saturate'' at
``small'' values because of the large number of fields entering the loops
 (e.g. the large number $N$ of gauge bosons, 
or the large value of the quadratic Casimir $C_A$,
for gauge groups like $E_8$). Typically, eq. (\ref{1}) would be replaced
(at $\phi \gg1$) by
\beq
\alpha_{GUT} ^{-1} \sim C_A +
{\cal O}(e^{-\phi}) ~~~, ~~ (M_P/M_s)^2 \sim N +
{\cal O}(e^{-\phi}) .
\eeq
In this picture there is naturally an asymptotic decoupling
mechanism of ordinary matter to the dilaton, whose effective mass
goes to zero at late times.  The problem remains, of course, of explaining
why the cosmological constant vanishes in superstring/M-theory, not
only at zero coupling where supersymmetry protects it, but also at
infinite (bare) coupling. Possibly, some new, stringy symmetry can
explain this. It will simply be assumed to be the case in  this paper.

As the dilaton is
non-universally coupled to different types of matter fields, its
coupling  to ordinary  matter can be asymptotically tiny (as to satisfy  
constraints from gravitational experiments \cite{16}), and much 
stronger (as first suggested  in \cite{damourgib}) to typical dark matter
candidates, such as the axion. In that case, the dilaton to
dark-matter  coupling leads  to an initial evolution, which is
similar to  the ``tracking" regime \cite{15}  of
conventional models of quintessence, but  takes place {\it before}
potential energy becomes appreciable. Later on, the interplay of the
dark-matter dilatonic charge and of the dilaton potential  leads to an
accelerated expansion in which the relative fraction of dark energy and
dark matter remains fixed (and of order 1),  thus offering a possible
explanation of the cosmic coincidence, as we will 
illustrate through explicit examples. 

The paper is organized as follows. In Sect. \ref{II} we present the
effective string cosmology equations, in the  small curvature --but 
arbitrary coupling-- regime, with generic matter sources non-minimally
coupled to the dilaton. In Sect. \ref{III} we discuss analytically
a possibile late-time attractor characterized by a constant positive
acceleration and a fixed ratio of dark matter and dark energy. In Sect.
\ref{IV} we provide a semi-quantitative description of the previous
phase, during which the dilaton potential can be neglected. This phase is
characterized by a ``focusing'' of the energy densities of the various
components of the cosmological fluid (which occurs before the epoch of
matter--radiation equilibrium), and by a subsequent ``dragging'' regime
in which the dilaton energy density tends to follow that of
non-relativistic (dark) matter. We also discuss here the main
phenomenological constraints that have to be imposed on the scenario.
In Sect. \ref{V} we consider a typical example of string cosmology
model  including radiation, baryonic and cold dark matter, and we
present the results of explicit numerical integrations. Our conclusions
are  summarized  in Sect. \ref{VI}.

 \section {Cosmological equations in the String and Einstein frames} 
\label{II}

Our starting point is the string-frame, low-energy, gravi-dilaton 
effective action \cite{10}, to lowest order in the $\a'$ expansion, but
including dilaton-dependent loop (and non-perturbative) corrections,
encoded in a few  ``form factors"  $\psi(\phi)$,  $Z(\phi)$,
$\alpha{(\phi)}$, $\dots$, and in an effective dilaton potential $V(\phi)$
(see also \cite{16}).
In formulae:
\bea
S &=& -{M_s^{2}\over 2} \int d^{4}x \sqrt{- \wt g}~
\left[e^{-\psi(\phi)}\widetilde R+ Z(\phi)
\left(\wt\nabla \phi\right)^2 + {2\over M_s^{2}} V(\phi)\right] 
\nonumber \\
&-& {1\over 16 \pi} \int d^{4}x {\sqrt{- \wt g}~  \over  \alpha{(\phi)}}
F_{{\mu\nu}}^{2} + \Ga_{m} (\phi, \wt g, \rm{matter})  
\label{3}
\eea
(conventions: metric signature: $(+,-,-,-)$,
$R_{\mu\nu\a}\,^\b= \pa_\mu \Ga_{\nu\a}\,^\b - $ \dots, $R_{\mu\nu}=
R_{\a\mu\nu}\,^\a$). 
Here $M_s^{-1} = \la_s$ is the fundamental string-length parameter,
$\wt g$ is the sigma-model metric minimally coupled to fundamental
strings,  $\wt R, \wt \nabla$ are the curvature and the covariant
derivative referred to $\wt g$, and $F_{\mu\nu}$ is the gauge field of
some fundamental (GUT) group ($\a(\phi)$ is the corresponding gauge
coupling). We imagine  having already compactified $6$ dimensions and 
having frozen the corresponding moduli at the string scale. Following
the basic proposal made in \cite{13}, we shall assume  that the
form factors $\psi(\phi)$,  $Z(\phi)$, $\alpha{(\phi)}$ approach a finite, 
physically interesting limit as $\phi \rightarrow + \infty$ while, in the
same limit,  $V \rightarrow 0$.

The fields appearing in the matter action $\Ga_{m}$ are
in general non-minimally and non-universally coupled to the dilaton 
(also because of the loop corrections \cite{17}). Their  
gravitational and dilatonic ``charge densities", $\wt T_{\mu\nu}$ and
$\wt \sg$, are defined as follows:
\beq
{\da \Ga \over \da \wt g^{\mu\nu}} ={1\over 2} \sqrt {-\wt g}~
 \wt T_{\mu\nu}, ~~~~~~~~~~~
{\da \Ga \over \da \phi} = - {1\over 2}\sqrt {-\wt g}~
\wt \sg, 
\label{4}
\eeq
and it is important to stress that, when $\wt \sg\not= 0$, the
gravi-dilaton effective theory is radically different from a typical,
Jordan--Brans--Dicke type model of scalar-tensor gravity \cite{18}. We
shall give a prototype form of $\Ga_{m}$  in the following section, after
passing to the Einstein frame.

The variation of (\ref{3}) with respect to $\wt g_{\mu\nu}$ then gives 
the equations:
\bea
&&
\wt G_{\mu \nu} + 
\psi' \, \wt {\nabla}_\mu \wt{\nabla}_\nu \phi   + \left[\epsi  Z - 
{\psi'}^{\, 2} + \psi''\right]  
\wt \nabla_\mu \phi \wt{\nabla}_\nu \phi  
\nonumber \\
&&  
+\frac{1}{2} \wt{g}_{\mu \nu}\left[\left(2 {\psi'}^{\, 2} - 2 \psi'' - \epsi 
Z\right)  (\wt{\nabla} \phi)^2
- 2 \psi' (\wt{\nabla}^2 \phi) - \epsi V(\phi)\right]\ =
\lambda_s^2 \epsi \wt {T}_{\mu \nu} ,
\label{5}
\eea
where $\wt G_{\mu \nu}$ is the Einstein tensor, and a prime denotes
differentiation with respect to $\phi$. The variation with respect to
$\phi$, using the trace of eq. (\ref{5}) to eliminate $\wt R$, leads to the
equation
\bea
&&
\left(3 {\psi'}^{\, 2} - 2 \epsi  Z\right) (\wt{\nabla}^2 \phi) +
\left[\epsi \left( Z \psi' -
 Z'\right) + \psi' \left(3 \psi'' - 3 {\psi'}^{\, 2}\right)\right] (\wt
\nabla \phi)^2 \nonumber \\  
&&
+ \epsi\left(2 \psi' V + V'\right) + \lambda_s^2 \epsi \left( \psi' \wt{T} 
 +  \wt{\sg}\right) = 0
\label{6}
\eea

We shall assume an isotropic, spatially flat metric background
(appropriate to the present cosmological configuration), and a perfect
fluid model of source. In the cosmic-time gauge we thus set 
\beq
\wt g_{\mu\nu}= {\rm diag} \left( 1, -\wt a ^2(\wt t)~ \da_{ij}\right), 
~~~~~~ \wt T_\mu^\nu = {\rm diag} \left (\wt\r, -\wt p
~\da_i^j \right),  ~~~~~~ \phi=\phi(\wt t), ~~~~~~~ \wt \sg = \wt \sg
(\wt t), 
\label{7}
\eeq
and one can easily check, combining the above equations, that the
matter stress tensor is not covariantly conserved (even in this frame),
but satisfies the equation
\beq
\dot{\wt{\rho}}+3 \wt{H}(\wt{\rho} + \wt{p}) = \frac{\wt \sg}{2} 
\, \dot \phi.
\label{8}
\eeq

For the purpose of this paper, and for an easier comparison with
previous discussions of the quintessential scenario, it is however
convenient to represent the dynamical evolution of the background in
the more conventional Einstein frame, characterized by a metric
$g_{\mu\nu}$ minimally coupled to the dilaton, and defined by the
conformal transformation $\wt g_{\mu\nu}=c_1^2 g_{\mu\nu}\epsi$.
Here $c_1^2$ parametrizes the asymptotic behaviour of $\psi(\phi)$,
\beq
\ c_1^{\, 2}=
\lim_{\phi \rightarrow + \infty} \exp \{-\psi(\phi)\},
\label{9}
\eeq
and thus controls  the asymptotic ratio between the string and the 
Planck scale, $M_P^2 = c_1^2 M_s^2$. In
 the Einstein frame the action (\ref{3}) becomes:
\bea 
S &=& -{M_P^{2}\over 2}\int d^{4}x \sqrt{- g}~
\left[ R - {k(\phi)^{2}\over 2} 
\left(\nabla \phi\right)^2  + { 2\over M_P^{2}}\hat{V}(\phi)\right]
\nonumber\\ 
&-&   {1\over 16 \pi} \int d^{4}x {\sqrt{- g}~  \over  \alpha{(\phi)}}
F_{{\mu\nu}}^{2} + \Ga_{m} (\phi, c_1^2 g_{\mu\nu}\epsi, \rm{matter})
\;,  
\label{EFaction}
\eea
where we have defined
\beq
k^2(\phi) = 3 \psi^{\prime 2} - 2 \epsi Z , 
~~~~~~~~~~~~~\hat V = c_1^4 e^{2\psi} 
V \; .
\eeq
For later use it is also convenient to define a canonical
dilaton field by:
\beq \label{canonic}
d \hat{\phi} = {M_{P} \over \sqrt{2}} k(\phi) d \phi \; ,
\eeq
although, in solving the equations, it will be easier to work directly 
with the original field $\phi$.

We now choose, also in the Einstein frame,  the
cosmic-time gauge, according to the rescaling
\bea
&&
\wt a = c_1 a e^{\psi/2}, ~~~~~~~~~
d\wt t = c_1 dt e^{\psi/2}, \nonumber \\
&&
\r = c_1^2 e^{2\psi} \wt \r, ~~~~~~~~~~
p = c_1^2 e^{2\psi} \wt p, ~~~~~~~~~~
\sg = c_1^2 e^{2\psi} \wt \sg  ~~~~~~~~~~.
\label{10}
\eea
From the $(0,0)$ and $(i,j)$ components of eq. (\ref{5}) we obtain,
respectively, the Einstein cosmological equations
(in units such that $M_P^2 = c_1^2 M_s^2 \equiv (8\pi G)^{-1} = 2$)
\bea
&&
6H^2 = \r +\r_\phi, \label{11} \\
&&
4 \dot H + 6H^2 =-p -p_\phi,
\label{12}
\eea
while from the dilaton equation (\ref{6}) we get
\beq
k^2(\phi) \left(\ddot{\phi}+3 H \dot{\phi}\right) +
k(\phi)\, k'(\phi)\, \dot{\phi}^2 
+ \hat{V}'(\phi) + \frac{1}{2}\left[{\psi'(\phi)} (\rho - 3 p) + \sg
\right] = 0.
\label{13}
\eeq
In the above equations $H= \dot a /a$, a dot denotes differentiation
with respect to the Einstein cosmic time, and we have used the
definitions:
\beq
\r_\phi= {1\over2} k^2(\phi) \dot \phi^2 +\hat V(\phi), ~~~~~~~
p_\phi= {1\over2} k^2(\phi) \dot \phi^2 -\hat V(\phi). 
\label {14}
\eeq
The combination of equations (\ref{11})--(\ref{13}) leads finally 
to the coupled conservation equations for the matter and dilaton
energy density, respectively:
\bea
&&
\dot \r +3H(\r+p) -{1\over 2} \dot \phi \left[{\psi'(\phi)} (\rho - 3 p) + 
\sg \right]=0,
\label{15}\\
&&
\dot \r_\phi +3H(\r_\phi+p_\phi) +{1\over 2}  \dot \phi
\left[{\psi'(\phi)} (\rho - 3 p) + \sg \right]=0.
\label{16}
\eea

For further applications, and for a more transparent numerical
integration, it is also convenient to parametrize the time evolution of
all variables in terms of the logarithm of the scale factor,  $\chi
= \ln (a/a_i)$, where $a_i$ corresponds to the initial
scale\footnote{The relation between $\chi$ and the redshift $z$ is $\chi
= -\ln(1+z) + \ln(a_0/a_i)$, where $a_0$ is the present value of the
scale factor.}, and to separate the radiation, baryonic and
non-baryonic matter components of the cosmological fluid by setting 
\beq
\r=\r_r + \r_b + \r_d \equiv \r_r + \r_m , ~~~~~~~
p={1\over 3} \r_r, ~~~~~~~
\sg = \sg_r + \sg_b + \sg_d \equiv \sg_r + \sg_m.
\label{17}
\eeq
The dilaton equation and the Einstein equation (\ref{11}) can then be
written, respectively, as 
\bea
&&
2\, H^2 \, k^2 \, \ \frac{d^2\phi}{d\chi^2}\, +\, k^2 \left(\frac{1}{2}\rhm
+
\frac{1}{3}\r_r
+ \hat{V}\right)\frac{d \phi}{d\chi} \, +\, 
2 H^2\, k\, k'  \left(\frac{d\phi}{d\chi}\right)^2 + 2 \hat{V}' + \psi' \rhm \
+ \sg = 0, \label{18}\\
&&
 H^2 \left[6 -\frac{k^2}{2}\left(\frac{d\phi}{d\chi}\right)^2 \right] \,
\ =\ \rhm + \r_r + \hat{V}.
\label{19}
\eea
The matter evolution equation (\ref{15}) can be split into the various
components as  
\bea
&&
\frac{d \r_r}{d\chi} + 4 \r_r - \frac{\sg_r}{2} \frac{d \phi}{d\chi}  = 0,
\label{20}\\
&&
\frac{d \r_b}{d\chi} + 3 \r_b -\frac{1}{2}\left( \psi' \r_b + \sg_b \right)
\frac{d\phi}{d\chi}\ = \ 0.
\label{21}\\
&&
\frac{d \r_d}{d\chi} + 3 \r_d -\frac{1}{2}\left( \psi' \r_d + \sg_d \right)
\frac{d\phi}{d\chi}\ = \ 0.
\label{rhodark}
\eea

Finally, eq. (\ref{18}) is also equivalent to the dilaton conservation
equation (\ref{16}), which becomes
\beq
\frac{d \rho_\phi}{d\chi} + 6 \rho_\phi - 6 \hat{V}(\phi) + \frac{1}{2}
\left( \psi' \r_m + \sg \right) \frac{d\phi}{d\chi} \ = \ 0 .
\label{22}
\eeq

\section {Accelerated late-time attractors with 
constant  $\Omega_\phi$}
\label{III}

As a first step towards a ``dilatonic" interpretation of quintessence
we will now discuss the possibility that the above equations, 
together with a string-theory motivated potential and loop corrections,
are asymptotically solved by an accelerated expansion, $\ddot a >0$,
with frozen ratio $\r_m/\r_\phi$ of the order of unity. This last
property, in particular, is expected to solve (or at least  alleviate) the
cosmic coincidence problem  \cite{8,9}. 

Under the assumption made in \cite{13} that the form factors
appearing in (\ref{3}) have a finite limit as $\phi \rightarrow +\infty$,
and  assuming the validity of an asymptotic Taylor expansion, we write:
\bea
&&
e^{-\psi(\phi)}\, = \, c_1^2 + b_1 e^{-\phi} + {\cal O}(e^{-2\phi}),
\;~~~~~~~~~ Z(\phi)\, = \, - c_2^2 + b_2 e^{-\phi} +  {\cal
O}(e^{-2\phi})\; ,  \nonumber \\ 
&&
\alpha(\phi)^{-1} = \alpha_0^{-1} + b
e^{-\phi} + {\cal O}(e^{-2\phi}) \; , \label{23}
\eea
where $c_1^2, c_2^2$ are assumed to be of the same order (typically
of order $10^2$ since, as already noted, $c_1^2 =(M_P/M_s)^2$), and
$\alpha_0$ is to be identified with the unified gauge coupling at the GUT
scale, i.e. $ \alpha_0 \simeq 1/25$. 
Unlike the model discussed in \cite{18a} our model thus describes, in
the strong coupling limit $\phi \ra +\infty$, a minimally coupled,
canonical scalar field $\hat \phi= \sqrt{2} (c_2/c_1) \phi$, see eq.
(\ref{canonic}). In the opposite limit, $\phi \ra -\infty$, the
gravi-dilaton string effective action reduces, as usual, to an effective
Brans--Dicke model with parameter $\om=-1$. We note that it is not hard
to chose $\psi(\phi)$ and $Z(\phi)$ in such a way that the kinetic term of
the dilaton keeps the correct sign at all values of $\phi$ (see the
example given in Sect. \ref{V}). 

Similarly, the assumption that $V$ originates from
non-perturbative effects, and that $V \rightarrow 0$ as $\phi
\rightarrow  \infty$, allows us to write, quite generically:
\beq \label{asynpot}
\hat{V}(\phi)\, = \, V_0\, e^{-\phi} + {\cal O}(e^{-2\phi}).
\eeq
Since the overall normalization of the potential $V_0$ is
non-perturbative, it should be related  to the asymptotic value of the
gauge coupling $\alpha_0$ by
 an expression of the form:
\beq \label{minipot}
V_{0} = M_s^4 \, \exp \left(- {4  \over \beta \alpha_{0}}\right)
=M_{*}^4,
\eeq
with some model-dependent (one-loop) $\beta$-function coefficient
$\beta$.
 For a comparison with earlier studies of an exponential potential
\cite{19,20} we also note  that, when referred to the canonically
normalized dilaton field $\hat \phi$ defined in   (\ref{canonic}), the
Einstein frame potential (\ref{asynpot})   asymptotically exibits an
exponential behaviour $\hat V \sim \exp (-\la \hat {\phi}/M_{P})$, with
$\la = c_1/c_2=\sqrt 2/k$ at $\phi \ra +\infty$.

It is important to discuss the size of the potential needed for
the viability of our scenario. Since the acceleration of the Universe
appears to be a relatively  recent phenomenon (even, possibly, an
\emph{extremely} recent one, as recently  argued in \cite{turner}), the
potential $V$ must enter  the game very  late, i.e. at an
energy scale of the order of $\r^{1/4} \sim 10^{-3}$ eV.
Unless we want to play with an unnaturally large present
value of $\phi$, this also fixes the scale of the potential in
(\ref{asynpot})  as $V_0 \sim (10^{-3}{\rm  eV})^4$. As far as we know,
this is feature is common  to all quintessence  scenarios:
 the problem of an outstandingly small cosmological constant
is traded for  the introduction of another unnaturally small mass scale
$M_{*}$.

 In our context,  we easily find that, in order to have a properly
normalized potential, we need the constant $\beta$ appearing in the
exponent of (\ref{minipot}) to be somewhat smaller than the
coefficient $\beta_3$ of the QCD beta function (see also the discussion
after eq. (\ref{35a})), say $\beta \sim 0.6 \beta_3$. Given our ignorance
of the origin of the dilaton potential, this looks perfectly acceptable, a
priori.  However, this apparent resolution  of the fine-tuning problem
should not hide  the fact that the potential has to be adjusted very
precisely if one wants to start the acceleration of the Universe not
earlier that at red-shift $z \sim  {\cal O} (1)$, and not later than today.
To the best of our knowledge there is no obvious explanation, at
present, of this  aspect of the coincidence problem.

Let us now come to the matter sector of the action (\ref{EFaction}).
As a typical example of $\Ga_{m}$ we take:
\beq \label{matter}
\Ga_{m} (\phi,  g, {\rm matter}) = 
\int d^{4}x \sqrt{- g}~  \bar{N} \left[i {\partial} \!\!\!\slash+ 
m_{N}(\phi) \right] N   + {1\over2} \int d^{4}x \sqrt{-  g}~\left[ 
e^{\zeta(\phi)} 
(\partial_{\mu} D)^{2} - e^{\eta(\phi)} \mu^{2} D^{2} \right]
\eeq
the first term representing  baryonic matter, the second
(scalar) cold  dark matter, while the gauge term appearing explicitly in 
(\ref{EFaction}) can already represent the radiation component of the
cosmic fluid. 

The non-observation of appreciable cosmological variations
of the  coupling  constants  \cite{costanti}, as well as the precision
tests of Newtonian gravity \cite{Fis} in the context of long-range
dilatonic interactions, force us  to assume that ordinary matter and
radiation have nearly metric couplings to $\wt{g}_{\mu\nu}$,  i.e. that
$\sg_b,\, \sg_r \, \simeq \, 0$ as $\phi \ra \infty$. It is not hard to see
how such a near--vanishing of dilatonic charges can be achieved starting
from the actions (\ref{EFaction}), (\ref{matter}). Following ref. \cite{17}
we have: 
\beq \label{35a}
{\sg_b \over \rho_b} \sim {\partial \over \partial
\phi}\left({\rm ln} ~\Lambda_{QCD}\right), ~~~~~~~~~ 
{\sg_r \over \rho_r} \sim  {\partial \over \partial
\phi}\left( {\rm ln}~ \alpha  \right).
\eeq
 Given that $\Lambda_{QCD} \sim M_s \exp \left(- {1 /\beta_3
\alpha}\right)$ (with $\beta_3$ the coefficient of the QCD
$\beta$-function), and using (\ref{23}) for $\alpha$, it is clear that
both  $\sg_b$ and $\sg_r$ are exponentially suppressed at large,
positive $\phi$. This decoupling mechanism is similar in spirit to the one
proposed in \cite{16}, although it is supposed to occur at infinite bare
coupling.

In the dark matter sector, on the
contrary,  we shall assume more generic  quantum corrections. 
By taking for instance  the action in eq. (\ref{matter}), one has for
the dilatonic charge of dark matter: 
\beq \label{Qdark}
 \sg_d \ = \, \
- ~ \zeta '(\phi) \, e^{\zeta(\phi)} (\partial_\mu \, D)^2 + \eta'(\phi)  \, 
e^{\eta(\phi)} \mu ^2 D^2.
\eeq
Furthermore, the equations of motion for the $D$ field
give a relation between
the time-averaged quantities, $ e^{\zeta(\phi)} \langle \dot D ^2
\rangle\, = \mu^2 e^{\eta(\phi)} \langle D ^2 \rangle$ (which is
consistent with the interpretation of $D$  as  non-relativistic matter,
$p_d = 0$, as  assumed in the preceding section), and relate $\sg_d$ and
$\rho_d$ by a (generally $\phi$-dependent)  proportionality factor  
\beq
\label{26} \sg_d / \r_d \ \equiv \ q(\phi) \ = 
\ \eta '(\phi) - \zeta ' (\phi) .
\eeq

The lat--time behaviour  we will discuss takes place if we assume that,
in the  strong coupling limit (i.e., $\phi \gg1$), $q(\phi)$ tends to a
positive constant of order unity, and that the
dark matter component dominates over baryonic matter and radiation.
Thus, the regime we are  considering is characterized
(according to eqs. (\ref{23}), (\ref{asynpot})) by 
\beq
k^2(\phi) = 2c_2^2/c_1^2 = 2/\lambda^2,  \quad \sg  
= \sg_d,  \quad \r  = \r_d,  \quad q(\phi) = q = {\cal O}(1), 
\quad \sg_d = q ~ \r_d \, .
\eeq
It follows that the dilaton coupling to the stress tensor can be
asymptotically neglected with respect to the coupling  to the dilatonic
charge, as  $\psi' \simeq e^{-\phi}/c_1^2 \ll 1$. 
The dilaton and dark matter conservation equations (\ref{rhodark}), \,
(\ref{22})  and the Einstein equations   (\ref{11}),\, (\ref{12}) can then
be written, asymptotically, in the form
 \beq
\label{dark1} \dot \r_d +3H\r_d -\frac{q}{2}\, \r_m \dot \phi=0,
\quad \quad
\dot \r_\phi +6H\r_k +\frac{q}{2}\, \r_m\dot \phi=0,
\eeq
\beq
1=\Om_d+\Om_k+\Om_V, \quad \quad \quad
1+{2\dot H\over 3 H^2}=\Om_V-\Om_k,
\label{28}
\eeq
where we have defined 
\bea
&&
\r_d= 6H^2 \Om_d,
~~~~~~~~~~~~~~~~~~~~~\r_\phi=\r_k+\r_V,\nonumber\\ &&
\r_k= 6H^2 \Om_k= \dot \phi^2/ \lambda^2, ~~~~~~~~~~
\r_V= 6H^2 \Om_V=\hat V. ~~~~~~~~
\label {29}
\eea

We now look for solutions with asymptotically frozen dark-matter over
dark-energy ratio, and frozen ``equation of state". From the constraint
(\ref{28}) this is equivalent to the requirement that $\r_k$,
$\r_V$ and $\r_d$ scale in the same way, i.e. 
\beq
\frac{d \log\r_\phi}{d\chi}\, = \, \frac{d \log\r_d}{d\chi}, \quad \quad
\quad \frac{d \log\r_V}{d\chi}\, = \, \frac{d \log\r_d}{d\chi}.
\label{30}
\eeq
The first condition and the conservation equations give  
\beq
\frac{d \phi}{d \chi}\ =\ \frac{6}{q}\, (\Omega_V - \Omega_k)\, .
\label{31}
\eeq
Expressing $d\phi/d{\chi}$ through $\Om_k=(d\phi/d{\chi})^{2}
/6 \lambda^2$, and inserting it in the second condition
(\ref{30}), we obtain, respectively,
\beq
 \lambda ~ q =\sqrt{6 \over  \Om_k} \left(\Om_V-\Om_k\right),  \quad
\quad \quad 
 \label{dark-qd2}
q = 2\, \frac{\Omega_V - \Omega_k}{1 + \Omega_k - \Omega_V},
\eeq
where in the latter the asymptotic form of the potential (\ref{asynpot})
has been used.
The last two equations can be solved for $\Omega_k$ and $\Omega_V$, 
\beq
\Omega_k\, =\, \frac{6}{\lambda^2 (2+q)^2},
\quad \quad \Om_V\, =\, \Omega_k + \frac{q}{q+2}
\eeq
giving easily
\beq \label{33}
\Omega_\phi\, =\, \frac{12
+  q(q+2)\lambda ^2}{ (q+2)^2\lambda ^2}, \quad \quad \quad 
w_\phi\, = \, - \frac{ q (q+2)\lambda^2}{12
+  q(q+2)\lambda^2},
\eeq
where the last equation for $w_\phi= (\Om_k-\Om_V)/(\Om_k+\Om_V)$ 
provides the dilaton's 
equation of state.

The above asymptotic solution, first obained in \cite{30a}, and recently
studied in \cite{30b,30c}, generalizes the results discussed in
ref. \cite{19} to the interacting dark matter case, and
is very similar to the results obtained by including suitable
non-minimal couplings  in a Brans--Dicke context \cite{20}, or by
including an effective bulk viscosity in the dark matter stress tensor
\cite{8,Pavon}. Our eq. (\ref{dark1}) corresponds indeed, formally, to a
dissipative pressure $\Pi=- q\r_m (\dot \phi/6H)$ (in the notation of
\cite{Pavon}).  See also \cite{30b,30c} for a discussion of the
parameter values compatible with such an asymptotic solution. 

Once $\Om_k$ and $\Om_V$ are given, one can easily compute all the 
relevant kinematic properties of the asymptotic solution as a
function of only  two parameters, $q$ and $\la=c_1/c_2$,
which are in principle calculable for a given string theory model. The
asymptotic value of the acceleration, in particular, is fixed by eq.
(\ref{28}) as  
\beq
 {\ddot a \over aH^2}= 1+{\dot H\over H^2}= {q -1\over q+2}.
\label{35}
\eeq
One can also easily obtain, through a simple integration, the asymptotic
evolution of the Hubble factor and of the dominant energy density,
\beq
H \sim a^{-3/(2+q)}, ~~~~~~~~~~~~~~
\r \sim a^{-6/(2+q)}. 
\label{318}
\eeq

\begin{figure}[h]
\begin{center}
\includegraphics{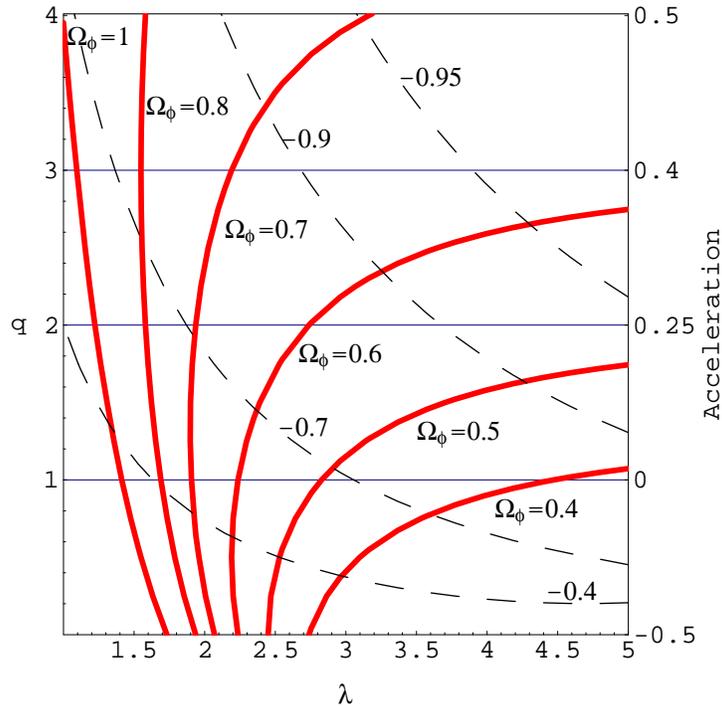}
\vskip 5mm
\caption{\sl 
The asymptotic configurations in the plane  $\{\la\ ,q\}$. 
The full bold curves correspond to asymptotic solutions with fixed ratios
$\r_\phi /\r_d$ and with the following values of $\Om_\phi $:
$ 1, \  0.8, \ 0.7, \ 0.6, \ 0.5, \ 0.4 $ .
On the right vertical axis we have reported the corresponding
$q$-dependent  acceleration parameter, $\ddot{a} a/\dot{a}^2$. The
thin dashed curves correspond to fixed asymptotic values of the
dilatonic equation of state $w_\phi= p_\phi/\r_\phi$, respectively
$-0.4$,  $-0.7$, $-0.9$ and $-0.95$. }    
\end{center}
\end{figure}

In order to illustrate the range of parameters possibly compatible with
present phenomenology, we have plotted in the $\{\la\ ,q \}$ plane
various curves at  $\Om_\phi= \Om_k+\Om_V=$ const, and $w_\phi=
(\Om_k-\Om_V)/(\Om_k+\Om_V)=$ const. (Fig. 1).
 Note that the case discussed in Ref. \cite{19} corresponds to
 staying on the $\lambda$ axis. In that case, 
the critical value of $\lambda$ below which 
$\Om_{d}/\Om_{{\phi}}\rightarrow 0$ is $\sqrt{3}$. The addition of  $q$
makes  parameter space two-dimensional,  with  the point $\lambda
= \sqrt{3}$ replaced by the left-most curve $\Om_\phi=1$.
Beyond that curve, i.e. for $\lambda^2 < 6/(2+q)$ (as well as for all
values of $\lambda$ if $q < -2$), the ratio
$\Om_{d}/\Om_{{\phi}}$ goes to zero. However, while in the case of
\cite{19} acceleration and a finite ratio $\Om_{d}/\Om_{{\phi}}$ are
incompatible, this is perfectly possible in a large region of the 
$\{\la\ ,q \}$ plane.

In fact, it is possible to determine the region of our
parameter space that survives the various
observational constraints (Type 1a supernovae, CMB anisotropies,
large-scale structure \dots). The present  values of   $\Om_\phi$ and
$w_\phi$  have to lie in the range \cite{Vincoli,32a} $0.6 \laq \Om_\phi
\laq 0.7$, and $-1 \leq w_\phi \laq -0.4$, but the two allowed intervals
are not uncorrelated. Assuming that we are already in the asymptotic
regime, the allowed region  lies roughly between the two curves
$ \Om_\phi = 0.6$ and $ \Om_\phi = 0.7$ and above $ q = 2$.  Other
phenomenological (but somewhat more 
model-dependent) constraints on $q$ and $\la$ can be 
obtained from the recent measurements of the position of the third
anisotropy peak in the CMB distribution \cite{21a}, which constrains the
value of $\Om_\phi$ today and at last scattering, as well as the
time-averaged equation of state $\langle w_\phi \rangle$ \cite{21b}. 
In the final part of this paper we shall present a model of dark
matter that seems to be compatible with all the above-mentioned
constraints.

\section {Focusing and dragging with {\bf $V=0$}: an analytic study of
 early-time evolution}
\label{IV}

Having discussed, in the previous section, the late-time accelerated
expansion caused by the interplay of the dilaton potential and the
dark-matter
dilatonic charge,  it looks appropriate to illustrate 
the  earlier evolution, i.e. {\it before} the dilaton potential
starts entering the game. In this section we shall provide a
semi-quantitative, analytic analysis of this behaviour as it follows
from the string cosmology equations (\ref{20})--(\ref{22}), 
by imposing on the non-perturbative normalization 
(\ref{minipot}) the constraint $V_0^{1/4} \ll H_{\rm eq}$, where 
$H_{\rm eq}$ is the curvature scale at the epoch of matter-radiation
equality. In such a way the dilaton potential may eventually become
important only at late times, in the matter-dominated era. 
We will show that this early  
evolution can be roughly divided in three epochs, 
providing, altogether, an intermediate attractor  
 that nicely connects to the accelerated behaviour
described in Section \ref{III}. 

Let us start by considering an initial, post-big bang and post-inflationary
regime of expansion
driven by the  standard radiation fluid, with negligible dilatonic charge,
$\sg_r=0$. Possible non-relativistic matter, if present, 
is highly subdominant with respect to the other components ($\r_m \ll
\r_\phi,\  \r_r$) and, consequently,
the  dilatonic terms in eq. (\ref{22}) can be neglected. 
The conservation equations can be easily integrated to give
\beq
\r_r=\r_{r i} e^{-4\chi}, ~~~~~~~~~~~~~
\r_\phi=\r_{\phi i} e^{-6\chi}.
\label{38}
\eeq
Therefore, the dilaton (kinetic)  energy density, even if initially of 
the same order as $\r_r$, is rapidly diluted like $a^{-6}$. 
The dilaton itself, starting from a value $\phi_i
\sim {\cal O} (1)$ typical of the moderately-strong coupling post-big
bang epoch, tends to settle down to a constant value (as already
noticed in \cite{34a}), that can be easily estimated as follows
 \beq
\r_k ={k^2\over 2}H^2 \left(d\phi\over d\chi\right)^2 =
{k^2\over 12}\left(d\phi\over d\chi\right)^2\left(\r_r+ \r_\phi 
\right)= \r_\phi.
\label{39}
\eeq
For $\r_{ri}=\r_{\phi i}$ we get 
\beq
{d\phi\over d\chi}= {\sqrt{12}\over k}\left (1 + e^{2\chi}\right)^{-1/2},
\label{40}
\eeq
which, for $k = $ const., leads to a solution with
 asymptotic value $\phi = \phi_1$, related to the initial value
$\phi_i=\phi(0)$ by the constant shift
\beq
\Da \phi= \phi_1-\phi_i= {\sqrt{12}\over k} \ln(1+\sqrt 2)
\simeq {3\over k} \simeq {3\over \sqrt 2} {c_1\over c_2}, 
\label{41}
\eeq
independently of $\phi_i$ and of the initial $\chi$ (the last equality
holds for $\phi_1$ large enough to justify the asymptotic relation 
$k=\sqrt 2/\la$). 

Such an initial regime is effective until the dilaton kinetic energy
becomes of the same order as $\r_m$. At that point, some
oscillations are triggered by the interference term  
of eq. (\ref{22}), but  the dilaton energy density keeps decreasing, on
the average, until it enters a ``focusing"  regime, during which it  is
diluted at a much slower rate (like $a^{-2}$), so as to converge, at
equality,  towards the larger values of $\r_m$ and $\r_r$. Eventually,
when dark non-relativistic matter becomes the dominant source ($\r_d
\gaq \r_r$), the dilaton energy density tends to follow the dark matter
evolution, as if it were ``dragged" by it. 

Before turning to a quantitative analysis of these two regimes we
note that the time evolution of $\r_\phi$, in the ``tracking"
quintessence, is
determined by the slope of the potential. In the present context,
instead, the focusing and dragging
effects are not due to the potential, but they are controlled by the
non-minimal coupling induced by $(\psi' + q)$ (thus implementing an
attractor mechanism already proposed for a class of non-minimal
scalar-tensor models of quintessence \cite{20a}). Thanks to the
focusing effect, which seems  to be typical of the string effective
action (even if similar, in a sense, to the ``self-adjusting" solutions of
general relativity with exponential potential \cite{19}), the dilaton
energy density at the  matter--radiation
equality turns out to be fixed
independently from its initial value, and only slightly dependent  from
the initial value of the dilaton, $\phi_i$. For large enough values of $q$,
however, even the dependence upon  $\phi_i$ tends to disappear,
because the value of the dilaton itself gets focused, as
will be discussed in the next section.  

For a quantitative analytical study of the ``focusing" and ``dragging"
regimes,  we start from eqs. (\ref{20})--(\ref{22}). Lumping
together baryonic  and dark matter, neglecting $V$,
and assuming, according to (\ref{26}),  $\sg = \sg_m = q(\phi) ~ \r_m$,
those equations can be easily recast in the form:
\beq
\r_r^{-1}\, \frac{d\r_r}{d \chi} + 4 = 0,
\eeq
\beq \label{gabrhom}
\r_m^{-1}\, \frac{d \r_m}{d \chi}  + \left[ 3  \mp \sqrt{3}\, \epsilon \, 
(\r_\phi/\r)^{1/2}
 \right] =0,
\eeq
\beq \label{gabrhophi}
\frac{d \r_\phi}{d \chi}  + 6 \r_\phi \pm  \sqrt{3}\, \epsilon\, \r_m 
(\r_\phi/\r)^{1/2} =0,
\eeq
where we have introduced the important parameter:
\beq \label{epsilon}
\epsilon(\phi) \, \equiv \, \frac{\psi'(\phi) + q(\phi)}{k(\phi)},
\eeq
and the sign ambiguity comes from solving eq. (\ref{39}) for
$d\phi/d{\chi}$ in terms of $\r_\phi$.
The focusing solution is then characterized by the relation:
\beq \label{ansatz}
\r_\phi \, = \, \frac{n^2(\phi)\, \r_m^2}{\r},
\eeq
i.e. $\Omega_\phi\, = \, n^2(\phi)\, \Omega_m^2$, which holds under the
assumption  that both $\epsilon$ and $n$ are slowly varying.
Indeed, we can establish the connection between these two quantities
by   inserting the ansatz (\ref{ansatz}) into (\ref{gabrhophi}). This 
gives:
\beq
-6 \mp \sqrt{3}\, \frac{\epsilon}{n}\, = \, 2 n ^{-1}\frac{d n}{d\chi} + 
2\r_m ^{-1} \frac{d\r_m}{d\chi} - \r ^{-1}\frac{d\r}{d\chi},
\eeq
where on the right--hand side the logarithmic derivative of
(\ref{ansatz}) has been taken. By using (\ref{gabrhom}) one finally has 
\beq \label{gabmain}
\r ^{-1} \frac{d \r}{d\chi}  \mp \sqrt{3}\, \epsilon \left[n^{-1} +
 2 \,\Om_m n \right]\, =\, 2 n^{-1} \frac{d n}{d\chi}\, \simeq\, 0.
\eeq

We can now discuss a few cases of interest.  During the
radiation--dominated 
phase, and after the kinetic energy of the dilaton is quickly
red-shifted away, we can neglect the term with $\Om_m$ in eq.
(\ref{gabmain}),  we set $d\r /d{\chi} =-4\r$, and obtain:
\beq \label{focusing}
n\, \simeq\, \frac{\sqrt{3}\, \epsilon}{4} , \quad \quad 
\frac{d\phi}{d \chi}\, \simeq \, -\frac{3\, \r_m \, \epsilon }{2\, k \, \r},
\quad \quad \r_\phi \, \simeq \, \frac{3\, \r_m^2 \, \epsilon^2 }{16\, \r}.
\eeq
We refer to this behaviour as ``focusing"  since it implies that $\r_m$
lies, modulo a factor $(16/3 )\epsilon ^{-2}$, at the geometric mean 
between $\r \sim \r_r$ and $\r_\phi$.
Hence, as we approach radiation-matter equality,  $\r_\phi$ is 
effectively focused towards the same common value of the other two 
components (see eq. (\ref{415}) below).
Note that, for a positive $\ep$, this happens thanks to a {\it negative}
$d\phi/d{\chi}$.

In the matter--dominated regime it is no longer safe to neglect 
the term in $\Om_m$ in eq. (\ref{gabmain}), unless $\ep \ll 1$. 
In that case, the  solution is
\beq \label{413}
n\, \simeq\, \sqrt{3}\, \epsilon \left(- \r^{-1}\frac{d
\r}{d\chi}\right)^{-1}, \quad \quad
\Omega_\phi\, \simeq \, 3\, \epsilon^2 \Omega_m^2 
\left(- \r^{-1}\frac{d \r}{d\chi}\right)^{-2}\; .
\eeq
During matter domination, using 
$d\r/d{\chi}=-3\r$, one gets 
\beq \label{materia}
n\, \simeq\, \frac{\epsilon}{\sqrt{3}}\, , \quad \quad 
\frac{d\phi}{d \chi}\, \simeq \, -\frac{2\, \epsilon }{ k },
\quad \quad 
\r_\phi\, \simeq \,\frac{\r_m \epsilon^2}{3}.
\eeq
In other words, the focusing regime has been turned into a
dragging one: the dilaton energy is dragged along by the (dark) matter
energy and  keeps a (small) constant ratio to it. Incidentally, at
the epoch of exact matter-radiation equality, using $d\r/d{\chi}
=-3.5\r$, we easily get (still at small $\ep$):
\beq
{\r_\phi\over \r_{eq}} \simeq \, \frac{3 \epsilon^2}{49}\; ,
\label{415}
\eeq
which is always smaller than $6 \% $ for $\ep < 1$.

In order to understand what happens at larger values of $\ep$ it
is useful to find the reason why, for small $\ep$,
$\r_\phi/\r_m$ stays constant. This comes about because the
corrections to the $a^{-3}$ and $a^{-6}$ laws for $\r_m$ and $\r_\phi$,
due to the non-vanishing $\ep$, push the two towards each other. It is
easy to check that, precisely if $\r_\phi/(\r_m + \r_\phi) = \ep^2 /3 $,
both energies scale like  $a^{-(3 +\ep^2)}$. We note, incidentally,  that
the above ratio of energies nicely fits with the value given in
(\ref{materia}) when $\ep\ll 1$. If $\ep < 1$,  the decrease of $\r_\phi $ 
is still slower than the $a^{-4}$ of $\r_r$, which justifies neglecting the
latter. However, if $\ep > 1$, this is no longer the case and we have a
third kind of behaviour, which can be called ``total dragging". In that
case, as shown by a simple analysis, all three components of $\r$ scale
like radiation, with the following sharing of the ``energy budget" 
(remember that we are always at $\Om = 1$): 
\beq \label{fulldrag}
\Om_\phi = \frac{\Om_m}{2} = {1 \over 3 \ep^2} , \quad \quad \Om_r = 
{\ep^2 -1 \over \ep^2}. 
\eeq
In the next section we will see how numerical integration
confirms in full detail the analytic behaviour we have discussed.
We end this Section by discussing 
some constraints on our parameters.

As already mentioned, we assume the ordinary  components of matter
(radiation and baryons)  to have a nearly metric coupling to ${\wt
g}_{\mu\nu}$ (see discussion after eq. (\ref{matter})). To be more
specific,  let us define the ratios between dilatonic charges and energy
densities in a way similar to that used  for cold dark matter in eq.
(\ref{26}), i.e.  
\beq \label{ordmag} 
q_r(\phi)\, \equiv \ \sg_r/\r_r , \quad \quad
q_b(\phi)\, \equiv \ \sg_b/\r_b . 
\eeq
Since it is  precisely the ratio $(\psi' + q_{r,b})/k$, which controls both
the effective coupling of the dilaton to ordinary macroscopic matter, as
well as a possible time-dependence of the fundamental
constants \cite{damour,dam99}, we shall  
assume that both $q_b$ and $q_r$ are at most of order  
$\psi'$, in agreement with the discussion after eq. (\ref{35a}). We then 
find that there are neither
appreciable violations of the equivalence principle in the context of
macroscopic gravitational interactions, nor significant contributions
to the time-variation of the fundamental constants, both
effects being controlled by $\psi'/k$ for $q_{r,b} \ra 0$.
In the strong coupling regime we have $\psi'/k \sim
e^{-\phi}/(c_1 c_2)$. For a non-negative $\phi_i$, and $c_1^2, c_2^2$ of
order $10^2$, there is no appreciable deviation from the standard 
cosmological scenario down to the epoch of matter-radiation
equality, so that one easily
satisfies the early-Universe constraints on dark energy, as reported for
instance in \cite{Melchiorri}.

The dilaton charge of dark matter is not restricted by the
experimental tests of long--range gravitational interactions:
this is the reason why we can play with it in order to produce an
acceleration.  Still, from the
above discussion on the early phases of the universe, it is clear that
high values of the dark-matter  parameter  $\epsilon$ may result in
dangerously high values for $\Omega_\phi$, and thus in  radical
deviations from the standard cosmological
scenario. Until radiation-matter equality the situation is relatively
harmless: we can easily estimate the
dilaton energy density at the equality and at the nucleosynthesis
scale,  $H_N \sim 10^{10} H_{\rm eq}$, using the fact that the dilaton,
during the focusing regime, is not significantly shifted away from the
value  $\phi_i +\Da \phi$, fixed by eq. (\ref{41}). 
Because of the focusing behaviour we find $\Om_\phi ({\rm
nucl})\sim 10^{-10} ~\Om_\phi ({\rm eq})$, and therefore
 the most stringent bound comes at equality, where, thanks to (\ref{415}),
 it is comfortably satisfied
for $\epsilon < 1$. 

During the dragging phase, however, we must certainly
impose $\epsilon < 1$, otherwise, the phenomenon of ``total
dragging''  takes place. This would represent a dramatic deviation
from the standard cosmological scenario, since all the components
$\r_\phi, \r_r, \r_d$ (except baryonic matter)  would redshift in the
same way ($a^{-4}$) from equality until  the   potential starts to be
felt. Even if $\epsilon < 1$,  but not sufficiently small, the
unusual scaling $\r_m \propto a^{-3 - \ep^2}$ tends to  change the
global temporal picture between now and the epoch of matter-radiation
equality and, from eqs.  (\ref{materia}), values of $\Omega_\phi
\sim \ep^2 /3$ (while in agreement with  possible constraints at last
scattering \cite{Melchiorri}) can be dangerously high. In our context,  a 
bound $\Omega_\phi({\rm drag}) < 0.1$, i.e. $\epsilon^2({\rm drag}) <
0.3$,  appears to be necessary in order to
agree with the observed CMB spectrum  and with the standard scenario
of  structure formation. 

On the other hand, due to the smallness of $\psi' \sim e^{-\phi}/c_1^2$
in the dragging regime,  an upper bound on $\epsilon$ effectively turns
into a bound on the value of  $q /k$, i.e. on the dilatonic charge of the
dark matter component. The above constraints thus translate into a
bound for the combination $\lambda q$:
 \beq
\lambda ~ q(\phi_{\rm drag})\, < \, 0.8 \, ,
\eeq
where we used the already mentioned asymptotic relation $\lambda =
\sqrt{2}/k$. It is clear that a constant $q$ cannot satisfy the above
bound and, at the same time,  provide the present acceleration of the
Universe by means of the mechanism described in Section \ref{III} (see
also Fig. 1), that requires $q \lambda \gaq 4$. 

A time-- (or, better,  $\phi$-- ) dependent $q$, however, is
 allowed. For this reason we have to
consider cold dark matter models like the one of eq. (\ref{matter}),
whose dilatonic charge (\ref{26}) switches on at large  enough values of
the dilaton. The transition to large values of $\phi$ is rapidly activated
as the potential comes into play, $\r_V\sim \r_\phi$. At that point, the
dilaton energy density stops decreasing and  freezes  at a constant
value,   necessarily crossing, at some later moment, the matter energy
density, $\r_\phi \sim \r_d$. From then on, the dilaton starts  rolling
towards $+ \infty$, triggering the effect of the dilatonic charge, which
rapidly freezes the ratio $\r_\phi/\r_m$ and (for suitable values of $q$)
leads to the accelerated asymptotic regime described by eqs.
(\ref{33})--(\ref{35}). Explicit numerical examples of such a behaviour
will be discussed in  Section \ref{V}.

For a realistic picture, in which the positive acceleration regime starts
around the present epoch (and not much earlier) and the
standard scenario of structure formation is implemented successfully, 
we have to require that the contribution of the dilatonic charge (as well
as the effect of the dilaton potential) come into play only at a late
enough epoch. The importance of this constraint was already discussed
in the context of other scalar-tensor models of quintessence \cite{20}
where, for instance, the non-minimal coupling of the scalar field to the
trace of the dark matter stress tensor was assumed to be
$\phi$-dependent, to interpolate between a small and a large mixing
regime.

\section {Numerical examples}
\label{V}

Finally, after the analytic discussion of the previous section, it seems 
appropriate to illustrate the ``run-away" dilaton scenario 
with some numerical example, both in order to confirm
the validity of some approximations made in deriving the
analytic results, and in order to see how the various regimes we
discussed can be put together.  To this aim, we shall 
numerically integrate eqs. (\ref{18})--(\ref{rhodark}), using eq.
(\ref{19}) as a constraint on the set of initial data, and assuming an
explicit model for the dilatonic charges and the dilaton potential. Also,
following the ``induced-gravity" ideas \cite{13}, we shall specialize the
loop form-factors according to eq. (\ref{23}), using the ``minimal"
choice
\beq 
e^{-\psi(\phi)}\, = \,  e^{-\phi} + c_1^2 \, , ~~~~~~~~~~~
Z(\phi)\, = \, e^{-\phi} - c_2^2 \, .
\label{51}
\eeq

First of all, for a clear illustration of the ``focusing" and ``dragging"
regimes, let us put $V=0$, $\sg_r=0=\sg_b$, and $\sg_d= q \r_d$, with
$q=$ const. By choosing, in particular, $c_1^2=100$, $c_2^2=30$, we
have integrated eqs. (\ref{18})--(\ref{rhodark}) for three different
values of the charge, $q=0$, $q=0.01$, and $q=0.1$, starting from the 
initial scale $H_i=10^{40}
H_{\rm eq}$, 
\beq
\left(a_i\over a_{\rm eq}\right)= \left(H_{\rm eq}\over H_i\right)^{1/2} =
\left(\r_{m i}\over\r_{r i}\right)= 10^{-20},
\label{52}
\eeq
and using $\r_{\phi i}= \r_{r i}$, $\phi_i=-2$ as initial conditions. It
should be noted that such initial conditions are generic, in the sense
that different initial values of $\r_\phi$ and $\phi$  may change the
fixed value reached by $\phi$ during the focusing phase, but do not
affect in a significant way the subsequent evolution, as will be
discussed at the end of this section. 

The results of this first numerical integration are illustrated in Fig. 2.
The left panel clearly displays the initial regime of fast dilaton dilution 
($\r_\phi \sim a^{-6}$), the subsequent focusing regime  ($\r_\phi \sim
a^{-2}$, see eq. (\ref{focusing})) triggered (after some oscillations) 
soon after $\r_\phi$ falls below $\r_m$, and the final dragging regime 
($\r_\phi \sim \r_m$, see eq. (\ref{materia})) in the epoch of matter
domination (the epoch of matter-radiation
equality corresponds to $\chi \simeq 46$).
Note that the constant values of $q$ have been chosen small enough to
avoid the phenomenon of ``total dragging", see Section \ref{IV}. Note
also that, in this example, $\r_m$ always coincides with $\r_d$. 
In the right panel the evolution of $\Om_\phi$, obtained through the
numerical integration, is compared with the analytic estimates
(\ref{focusing}), (\ref{413}), (\ref{materia}), for the three different
values of $q$. In all cases, $\Om_\phi$ grows like $a^2$ during the
focusing regime (in the radiation era), while the final stabilization
$\Om_\phi=$ const, after the epoch of matter-radiation
equality ($\chi \gaq 46$), clearly
illustrates the effect of the dragging phase during which $\r_\phi$ and
$\r_m$ evolve in time with the same behaviour. 

\begin{figure}[t] 
\begin{center}
\includegraphics{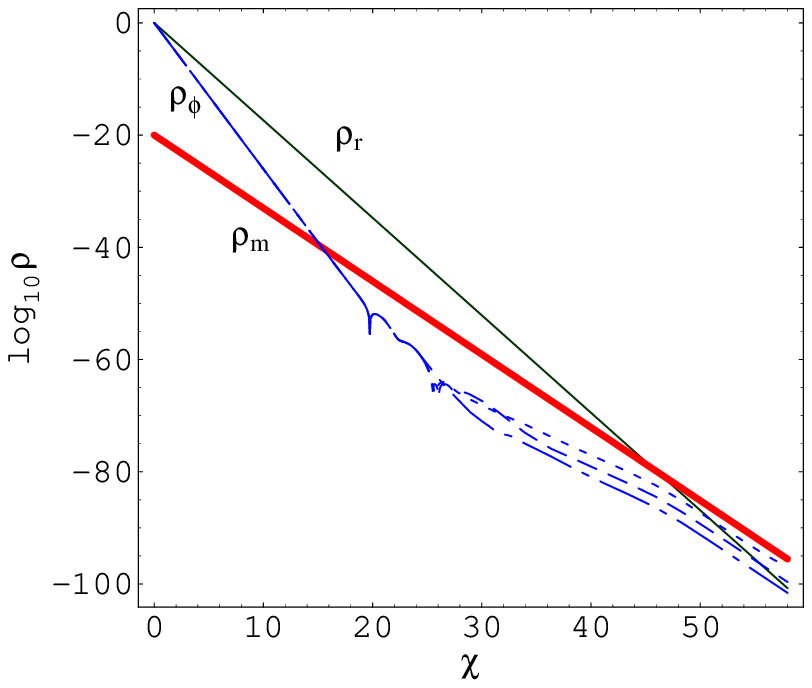}
\includegraphics{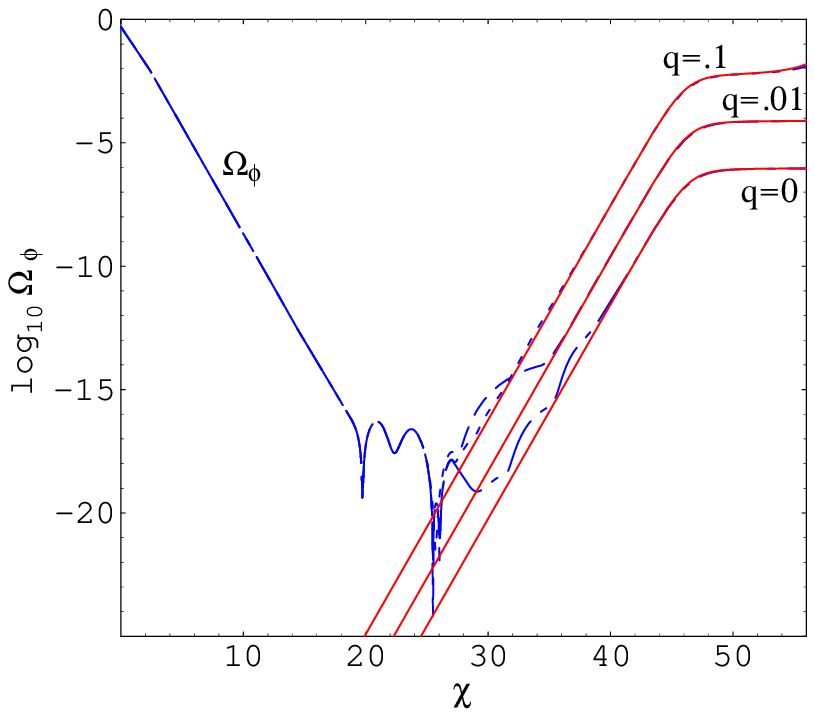}
\vskip 5mm
\caption{\sl Time evolution of $\r_\phi$ for $q=0$
(dash-dotted curve), $q=0.01$ (dashed curve) and $q=0.1$ (dotted
curve). The initial scale is $a_i= 10^{-20} a_{\rm eq}$, and 
the epoch of matter-radiation
equality corresponds
to $\chi \simeq 46$. Left panel: the dilaton energy density is compared
with the radiation (thin solid curve) and matter (bold solid curve) energy
density. Right panel: the dilaton energy density (in critical units) is
compared with the analytical estimates (\ref{focusing}), (\ref{413}),
(\ref{materia}) for the focusing and dragging phases.}
\end{center}
\end{figure}

For a realistic model of quintessence, however, a constant dilatonic
charge cannot drive the Universe towards an asymptotic accelerated
regime and, simultaneously, satisfy all the required phenomenological
constraints during the earlier epochs (as discussed in the previous
sections). By keeping $\sg_b$, $\sg_r \simeq 0$ at large coupling
 (see eq. (\ref{35a}) and the discussion thereafter), we shall
thus consider the explicit model of scalar dark matter (\ref{matter}),
with the following simple loop form--factors 
\beq
e^{- \zeta(\phi)} \, = \, 1 + e^{q_0 \phi}/c^2,
~~~~~~~~ e^{\eta(\phi)} = {\rm const}
\label{53}
\eeq
(note  that, by a field redefinition, one of the two loop factors can
always be taken to be trivial: what really matters is the
ratio $e^\zeta/ e^\eta$). Using (\ref{26}) we immediately get
\beq \label{54}
q(\phi)\, = {\sg_d \over \r_d}= q_0 \, {e^{q_0\phi} \over c^2 + e^{q_0
\phi}} \; , \eeq
which is exponentially suppressed in the perturbative regime, and
approaches $q=q_0$ at large coupling (for $q_0>1$ it is thus 
compatible with an asymptotically
accelerated cosmological configuration, see Fig. 1). 
For our numerical example we shall choose $q_0=2.5$
and $c^2=150$, but the behaviour of the solution is rather stable, at
late times, against large variations of the latter parameter (see the
discussion at the end of this section). 

In addition, we have to specify the form of the dilaton potential. In
agreement with its non-perturbative origin, and with the assumtion of
exponential suppression at strong coupling (see Section \ref{IV}), the
simplest choice is a  difference of  terms of the
type $e^{-\beta /\alpha(\phi)}$. We shall  thus consider the 
bell-like potential (in units $M_P^2=2$)
\beq
V(\phi)\ =\ m_V^{\, 2} \left[\exp\, (-e^{-\phi}/\beta_1) - 
\exp\, (-e^{-\phi}/\beta_2)\right],\quad \quad 0 < \beta_2 < \beta_1 ,
\label{55}
\eeq
which leads, asymptotically, to the large-$\phi$ behaviour of eq.
(\ref{asynpot}). The mass scale $m_V$, related to the mass $M_{*}$ of
eq.  (\ref{minipot}), will be fixed at $m_V=10^{-3}H_{\rm eq}$, together
with $\b_1=10, \b_2=5$, for a realistic scenario that starts
accelerating at a phenomenologically acceptable epoch. 

With all the parameters fixed, we have numerically integrated the
evolution equations (\ref{18})--(\ref{rhodark}), for our model of charge
(\ref{54}) and potential (\ref{55}), using the same initial conditions as
in the previous example, but separating  the
dark and baryonic components inside $\r_m$. In particular,
we have set, initially, $\r_{d i} =10^{-20} \r_{r i}$,  $\r_{b i} =7
\times 10^{-21} \r_{r i}$. 

The resulting late-time evolution of the various energy densities is
shown in the left panel of Fig. 3. Dark matter and baryonic energy
densities evolve in the same way, until the potential comes into play,
starting at a scale around $\chi \simeq 49$. The potential first tends to
stabilize $\r_\phi$ to a constant but then (thanks to the contribution of
$q$) the system eventually evolves towards a final regime in which
$\r_\phi$ and $\r_d$ are closely tied up, and their asymptotic evolution
departs from the trajectory of the standard, decelerated scenario (in
particular, they both scale, asymptotically as $a^{-6/(2+q_0)}$, see eq.
(\ref{318}). It is amusing to conjecture that the different
time-dependence of $\r_b$ and $r_d$ could be responsible for the
present small ratio $\r_b/\r_d$.

\begin{figure}[t]
\begin{center}
\includegraphics{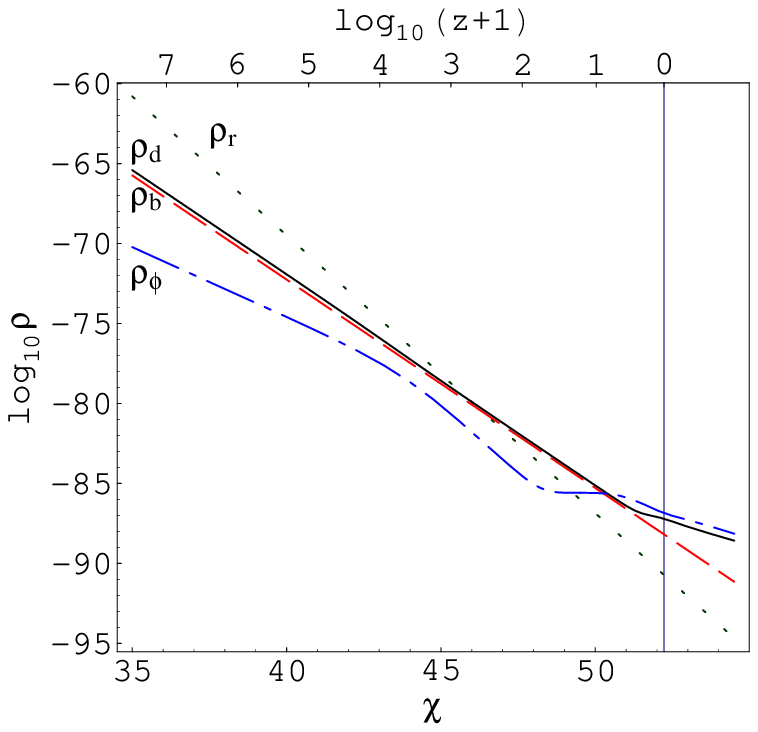}
\includegraphics{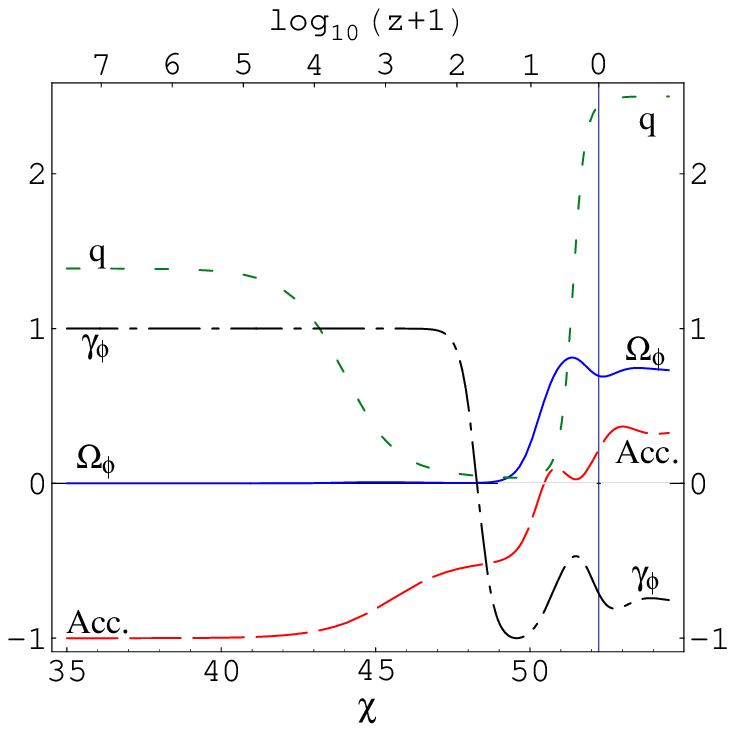}
\vskip 5mm
\caption{\sl Left panel: Late-time evolution of
the dark matter (solid curve), barionic matter (dashed curve), radiation
(dotted curve) and the dilaton (dash-dotted curve) energy densities,
for the string cosmology model specified by eqs. (\ref{54}), (\ref{55}).
The upper horizontal axis gives the $\log_{10}$ of the redshift
parameter.  Right panel: for the same model, the late-time evolution of 
$q$ (fine-dashed curve), $w_\phi$ (dash-dotted curve),  
$\Omega_\phi$ (solid curve) and of the acceleration
parameter $\ddot{a}a/\dot{a}^2$ (dashed curve).}
\end{center}
\end{figure}

In the right panel we have  plotted the time evolution of the dilatonic
charge $q$, of the energy density $\Om_\phi$, of the equation of state
$w_\phi$, and of the acceleration parameter $\ddot a/a H^2$. When
the potential energy becomes important, all the above quantities
rapidly approach their asymptotic values given in eqs.
(\ref{33})-(\ref{35}).  Note that, with our choice of parameters, we have
$q_0=2.5$ and $\la =c_1/c_2=\sqrt{10/3}$, corresponding  to an
asymptotic value $\Om_\phi \simeq 0.733$, slightly exceeding the best 
fit value suggested by present observations \cite{Vincoli,32a}. It is
important to stress, however, that the asymptotic attractor may be
preceded by a (short) oscillating regime, which, as illustrated in the
right panel of Fig. 3, can easily allow for values of the cosmological
parameters  different from the asymptotic ones to be compatible with
present observations. Note also that, when switching from the 
focusing to the dragging phases, the dilaton starts to move back
towards decreasing values of $q$, as will be illustrated also by a
subsequent numerical integration. This may slow down the evolution of
$\r_\phi$ with respect to $\r_m$ during the dragging, as shown for
instance in the left panel of Fig. 3. Because of this effect, however, the
dilaton  can easily satisfy, during the dragging phase, the
phenomenological bounds discussed in the previous sections. This does
not require fine tuning, the validity if the bounds being  guaranteed 
for a large basin of initial
conditions by  a convergent behaviour of the solutions during dragging.

\begin{figure}[t]
\begin{center}
\includegraphics{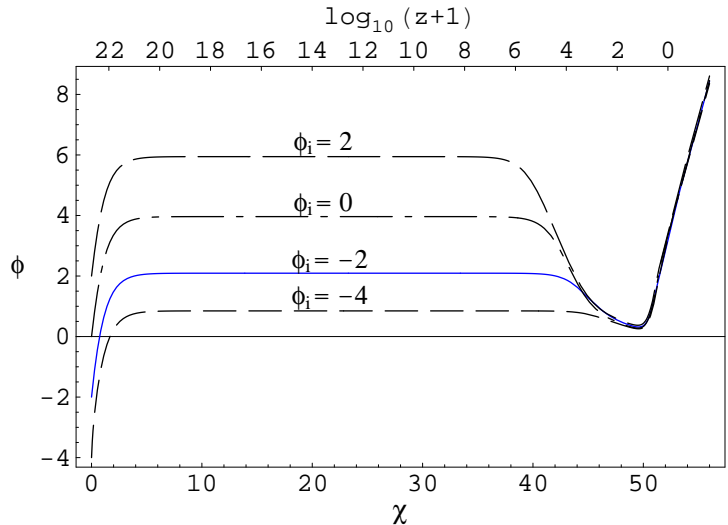}
\vskip 5mm
\caption{\sl Time evolution of the dilaton field, for different initial
conditions $\phi_i=-4,-2,0,2$. All the other parameters are the same as
in the example of Fig. 3. After the {\rm plateau} associated with the
focusing regime, and for a strong enough dilatonic charge, 
the solutions tend to converge to a common value of $\phi$.
The subsequent running to $+ \infty$, driven by the potential, is thus
completely independent from the initial value.}
\end{center}
\end{figure}

During the focusing phase, in fact, the dilaton is practically frozen, as
can be argued from eq. (\ref{focusing}), and   its effective
constant value, as determined by eq. 
(\ref{41}),  depends on $\phi_i$. However, if such a value is high enough, the
presence of the 
dilatonic charge may become important, and may contribute to the
focalization  towards the epoch of matter-radiation
equality, as already anticipated. 
This is illustrated in Fig. 4,  which shows the time evolution of
the dilaton obtained by numerically integrating the same model as in
Fig. 3, for different initial values $\phi_i=-4,-2,0,2$.
Although we start with different dilaton  values at 
the {\em plateau} associated with
the focusing regime, 
all the  solutions tend to converge as we enter the dragging regime, 
so as to make the 
subsequent (potential-dominated) evolution
{\em insensitive} to the initial value of the dilaton\footnote{The
\emph{preceding} evolution, of course, is not sensitive either, since  
during focusing the order of magnitude of $\Omega_\phi$ is given by
$\Omega_m^2$ as in (\ref{focusing}).}.

This new focusing effect, which is very different
from the one of the energy densities during the radiation-dominated 
phase, can also be understood analytically by writing the solution of 
eq. (\ref{materia}) as:
\beq
\chi - \chi_{\rm eq}\, =\, -\int_{\phi_{\rm eq}}^\phi
\frac{k(\bar{\phi})}{2\, \ep(\bar{\phi})}\, d \bar{\phi}\, .
\eeq
Since $k$ is almost constant,  a variation 
$\delta \phi_{\rm eq}$ on the initial value of $\phi$ changes 
the solution $\phi(\chi)$ by an amount $\da \phi(\chi)=[\epsilon (\phi)
/\epsilon(\phi_{\rm eq})]\delta \phi_{\rm eq}$, which rapidly
decreases (with $q(\phi)$) during the dragging phase. This is why the
solution has become independent of the initial value of  $\phi$ by the time
the potential becomes an important component. 

\begin{figure}[t]
\begin{center}
\includegraphics{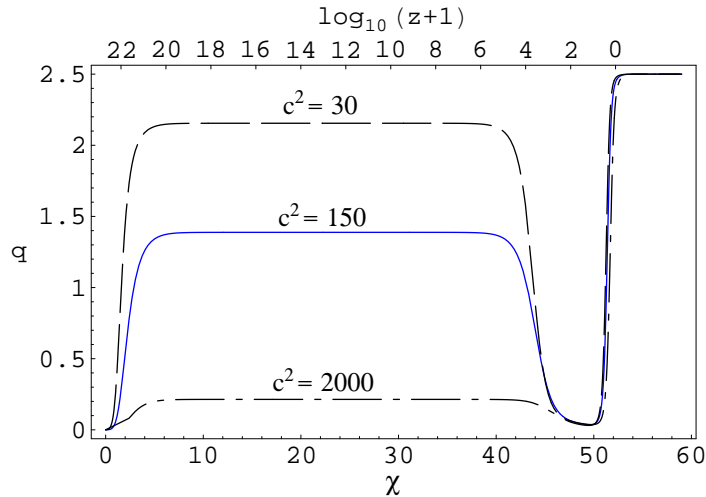}
\vskip 5mm
\caption{\sl Time evolution of $q(\phi)$, from eq. (\ref{54}), for three
different values of  the parameter $c$.  All the other parameters are the same
as 
in the example of Fig. 3. During the dragging phase the value of $q$
converges to the regime $q \ll 1$.}
\end{center}
\end{figure}

For the same reason, the model is only weakly affected by variations
of  the parameter $c$ in eq.
(\ref{54}), which roughly gives the transition scale 
between small and large dilatonic charges: $\phi_s = (2/q_0)\log c$.
Indeed, because of the above mechanism the dilaton is pushed  back 
during the dragging phase
the dilaton is pushed  back, with a velocity as high as needed  to
reach, in any case, the safe zone $q \ll 1$.
This effect is illustrated in Fig. 5, where we have plotted the time
evolution of $q(\phi)$, for the same model as Fig. 3, and for three
different values of $c$.

It should be noted, in conclusion, that the above class of models
depends in crucial way on three important parameters: $m_V$, $q_0$ and
the ratio $\la=c_1/c_2$. The first one controls the transition time
between the epoch of standard cosmological evolution and the final
accelerated regime (as can be easily checked, for instance, by
repeating the numerical integration of Fig. 3 with different values of
$m_V$). The other two parameters control the asymptotic properties of
the model (acceleration, equation of state, \dots), as discussed in
Section \ref{III}. Future precision data, both from supernovae
observations and from measurements of the CMB anisotropy, could
give us a good determination of these parameters, thus providing
important indirect information on the parameters of the string
effective action in the strong coupling regime.

\section {Conclusion}
\label{VI}

Let us conclude by summarizing the main points of our work.
We have argued that a run-away dilaton can provide an
interesting model of quintessence under a well-defined set of
assumptions that we list hereafter:
\begin{itemize}
\item The limit of supertring theory,  as its bare four-dimensional 
effective coupling goes to infinity (so called induced gravity/gauge or
compositeness limit), should exist and make sense phenomenologically,
i.e. should yield  reasonable values for the unified gauge coupling at the
GUT scale and for the ratio $M_P/M_{GUT}$, thanks to the large number of
degrees of freedom at $M_{GUT}$;

\item In the visible-matter sector, the couplings to the dilaton,
either direct or through the trace of the energy-momentum tensor (i.e.
via a conformally rescaled metric), should vanish in the $\phi \ra +
\infty$ limit;

\item In the dark matter sector, there should be a surviving coupling
to the dilaton (and thus violations of the strong and/or weak
equivalence principles) even in the $\phi \ra + \infty$ limit;

\item The dilaton potential should be non-perturbative, go to zero
asymptotically, and have an absolute scale not too far from the
present
energy density.
\end{itemize}

Under these circumstances, it is natural for the dilaton energy in
critical units, $\Om_{\phi}$, to be: i) subdominant  during radiation
domination; ii) a (small) fraction of the total energy at
matter-radiation equality; iii) a (small) fraction of $\Om_m$ during
the earlier epoch of matter domination; iv) a fraction of dark-matter
energy since a red-shift ${\cal O}(1)$. This very last phase is
characterized by an accelerated expansion.

In other words, this framework seems to be naturally consistent with
present astrophysical observations and with known cosmological
constraints.  From a theoretical  point of view the model appears to
combine nicely previous ideas  \cite{16} on dilaton stabilization and
decoupling with those recently advocated (e.g. in \cite{20}) so as to
have acceleration while keeping the ratio
$\Omega(\rm{dark\;energy})/\Omega(\rm{dark\;matter})$
constant.  

It must be stressed, of course,  that the analysis
presented in this paper is still preliminary, and that various problems
are still open. In particular, a precise computation of the CMB
anisotropy spectrum, and of the spacing of acoustic peaks
expected in this context, as well as a comparison with currently available
measurements \cite{4}, could provide significant bounds on the
parameters of the  string cosmology models we have discussed. Such an
investigation  is  postponed to  future work.  
Nevertheless, we believe that the results of this paper are
encouraging, as they suggest that the dilaton, which can hardly play
the role of the inflaton in the standard inflationary scenario \cite{21},
could play instead a successful role as the quintessential field in 
post-inflationary, late-time cosmology. 

 \acknowledgements
It is a pleasure to thank  Luca Amendola, Thibault Damour, 
Michael Joyce, Alessandro
Melchiorri, and J. P. Uzan for useful discussions. 
G.V. wishes to acknowledge the support of a ``Chaire Internationale  
Blaise Pascal", administered by the ``Fondation de L'Ecole Normale  
Sup\'erieure'', during most of this work.
F.P. would like to thank Luciano Girardello for his kind support,
the ``Centre Francais pour l'accueil et les \'echanges internationaux" 
for a scholarship, and  the ``Laboratoire de Physique Th\'eorique,
Universit\'e Paris   Sud,  Orsay" for hospitality.

\end{document}